\RequirePackage{fix-cm}
\documentclass[smallextended]{svjour3}       
\smartqed  

\usepackage{lineno} 
\usepackage[ruled]{algorithm2e}
\usepackage{amsmath} 
\usepackage{amssymb}
\usepackage{csquotes} 
\usepackage{graphicx}

\begin{document}
	
	\title{Dynamic Discovery of Type Classes and Relations in Semantic Web Data }
	
	\author{Serkan Ayvaz         \and
		Mehmet Aydar
	}

	\institute{Serkan Ayvaz \at
		Department of Software Engineering, Bahcesehir University, Besiktas 34353, Istanbul, Turkey.\\
		\email{serkan.ayvaz@eng.bau.edu.tr}           
		\and
		Mehmet Aydar \at
		Department of Computer Science, Kent State University, Kent, Ohio 44240, USA.\\
		\email{maydar@kent.edu}  
	}
	

	\maketitle
	
	\begin{abstract}
		The continuing development of Semantic Web technologies and the increasing user adoption in the recent years have accelerated the progress incorporating explicit semantics with data on the Web. With the rapidly growing RDF (Resource Description Framework) data on the Semantic Web, processing large semantic graph data have become more challenging. 
		Constructing a summary graph structure from the raw RDF can help obtain semantic type relations and reduce the computational complexity for graph processing purposes. 
		In this paper, we addressed the problem of graph summarization in RDF graphs, and we proposed an approach for building summary graph structures automatically from RDF graph data. Moreover, we introduced a measure to help discover optimum class dissimilarity thresholds and an effective method to discover the type classes automatically. 
		In future work, we plan to investigate further improvement options on the scalability of the proposed method.
		\keywords{Semantic Web \and RDF \and Graph Summarization \and Automatic Weight Generation}
	\end{abstract}

	\section{Introduction} 
	The Web as the global information source is growing exponentially.
	In recent years, there has been significant developments in publishing data with expressed semantics in the Web.  The Linking Open Data\cite{bizer_linked_2009} and similar community projects have recommended the publication of large amount of globally useful datasets in machine-readable forms. Moreover, the utilization of Resource Description Framework (RDF), along with other forms of semantic data in forms of RDFa \cite{adida_rdfa_2008} and microformats \cite{khare_microformats:_2006} in web pages has expanded.
	
	As a standard data model for the Semantic Web, RDF is a graph-structured general purpose language for representing information in a way that the resources are described unambiguously using RDF statements. The RDF statements are in the form of subject-predicate-object triples. Every triple is a relationship between two entities.
	
	RDF uses rdf:type property for stating class membership of entities. The entity type information is particularly useful for semantic searches in finding related entities and in traversal of the hierarchical structure of the RDF graph. However, the semantic data available on the Web today often don't have precise entity type information. It is partially due to (1) not containing the entity types owing to the flexibility of RDF model not forcing constraints on the schema, (2) representing data with non-standard vocabularies for typing 
	as some data publishers do not use standard vocabularies such as rdf:type and rdfs:subClassOf, 
	which is making it challenging to locate the type triples, furthermore, (3) defining the type information too generally that loosely coupled entities are represented in the same types.
	
	Constructing a graph structure containing the entity type classes, class attributes and relations between the type classes can be instrumental for Semantic Search algorithms in terms of query time since the entire input data does not need to be completely processed at the query time. We call this structure as the Summary Graph \cite{ayvaz_building_2015,aydar_automatic_weight}.
	Semantic search is a common graph processing task and it often requires a summary graph structure for effective and faster graph processing. For instance, the semantic search approach proposed by \cite{tran2009top} uses a summary graph structure in the search mechanism. They generate the summary graph using a set of aggregation rules, which calculate the equivalence classes of all nodes belonging to one type class and project all edges to corresponding edges accordingly. In this approach, one needs to know what constitutes a type class in advance. However, this assumption may not be realistic for real-world RDF datasets, i.e., the RDF data may not be tied to a standard ontology or vocabulary. Our work attempts to address this issue by automatically building the summary graph structure from the data itself by utilizing graph node similarity scores.
	
	There exists related methods to obtain a summary graph: (1) A summary graph can be obtained from the dataset ontology, if the dataset is already tied to an ontology. (2) Another way to obtain the summary graph is to locate the type triples (rdf:type) in the dataset and to organize the type classes and relations accordingly, if the data set is published using a standard vocabulary \cite{duan2011apples}. (3) Or the summary graph can be built automatically without relying on an  ontology or a standard vocabulary.
	Our graph summarization approach is based on the latter method. 
	Rdf:type is an optional property and it is often missing in commonly used datasets. Furthermore, it can potentially be inconsistent or erroneous in some cases \cite{parundekar2012discovering}. Thus, the methods, which rely on rdf:type or existence of an ontology may have implications for general use. Therefore, there is a need for automatic generation of summary graphs from RDF Data. In this regard, we describe an entity similarity metric and the methods used for automatically generating a summary graph from RDF Data. To the best of our knowledge, this is the first approach to attempt to generate summary graph of RDF graph automatically based on the entity similarities.

	\subsection*{Contributions and Outline}
	In this study, we focus on the problem of efficiently building a summary graph structure automatically from underlying RDF data. We utilize the notion of entity similarity in an RDF dataset so that fundamentally similar entities could be associated with the same class, which we call it type class in this paper.
	
	In our approach, the type classes, importance weight of each property and each string word for each of the referenced IRIs (Internationalized Resource Identifiers) are auto-generated. Furthermore, the weights in the pairwise similarity calculation are generated dynamically and applied during the summary graph generation. 
	Our methodology is to utilize graph locality and neighborhood similarity. Our algorithm does not rely on the existence of a common vocabulary. We use the Jaccard measure context for entity similarity such that the properties of the entities are treated as the dimensions of the entities.
	
	A stability measure, which represents the degree of confidence of a relation between classes in the summary graph, is proposed.
	From the input RDF data itself, we generate the summary graph along with the classes and class relations with the stability measure for each class relation.
	The main contributions of present study are
	\begin{itemize}	
		\item  We investigated the graph summary problem in RDF graphs and provided an effective approach for generating summary graphs automatically from RDF data. 
		\item For automatic discovery of the summary graphs, we introduced a measure, which we call Favorability that helps discover optimal class dissimilarity thresholds and provided an effective method that discovers the type classes using the class threshold measure. 
		\item To assess the effectiveness of our approach, we applied our methods to real-world datasets.
		
	\end{itemize}
	
	The rest of the paper is organized as follows.  
	In section \ref{Defining Graph Summarization Problem}, we define the graph summarization problem in RDF graph data. Then, in section \ref{Methods}, we discuss  our methods and the algorithms in detail. In section \ref{Evaluations}, the results of the evaluations assessing the efficiency of the proposed methods are presented. Finally, we review the related work in section \ref{Related Work} and follow this with our conclusion and future work in section \ref{Conclusion}.

	\section{Defining Graph Summarization Problem}\label{Defining Graph Summarization Problem}
	
	RDF data consist of a collection of statements that intrinsically represent a labeled, directed multi-graph with which the resources are expressed unambiguously. RDF statements describe resources in the form of triples, consisting of subject-predicate-object expressions that describe a resource, the type of a resource (type triple), or a relationship between two resources \cite{cyganiak_rdf_2014}. 
	
\begin{figure}[htbp]
		\centering
		\includegraphics[height=9.5 cm]{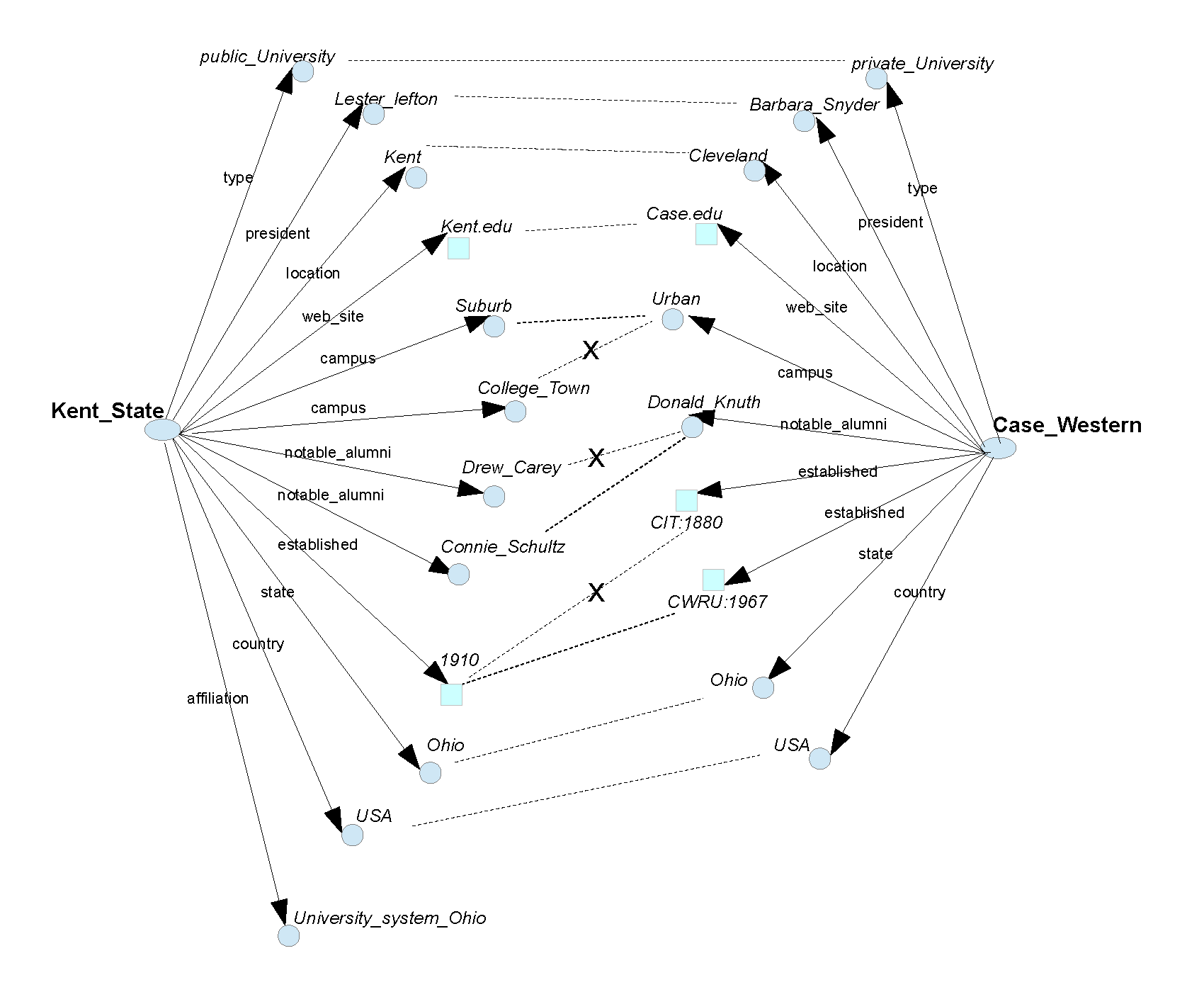}
		\caption{Sample graph demonstrating two nodes}
		\label{fig:figureKent}
	\end{figure}

	To describe resources, each RDF node that corresponds to an RDF entity is represented with an IRI. The values such as strings, numbers and dates in RDF data are represented by literal nodes.  A predicate in an RDF triple is also called a property of the RDF subject node. A predicate can be one of two types: a DatatypeProperty where the subject of the triple is an IRI and the object of the triple is a literal or an ObjectProperty where both the subject and object of the triple are IRIs. Each object of a subject node is called a neighbor of that subject node. 
	The subject in an RDF triple is either an IRI or a blank node, the predicate is an IRI, and the object is either an IRI, a literal or a blank node. The subjects and objects of triples in the RDF graph form RDF nodes. As an example, figure \ref{fig:figureKent} represents two sample university entities $Kent\_State$ and $Case\_Western$ and their properties.

	Formally, RDF graph is a directed labeled graph, which can be represented as $G= (V,L,E)$, where the set of vertices $V$ represents entities (resources), the set of directed edges $E$ of the form $l(u, v)$,  with  $u\in V$,  $v\in V$ and  $l \in L$, denote predicates (properties) between entities, and the labels $L$ are predicate names or labels. Note that an edge $l(u,v)$ represents the RDF triple $(u,l,v)$.

	A summary graph of a data graph is the directed graph such that each node in the summary graph is a subset of the original graph nodes of the same type. Thus, we define the Graph Summarization Problem as finding the corresponding summary graph $G'= (V',L',E')$ of $G$, 
	such that each element of $V'$ is a subset of $V$ containing all elements of the same type.
	For $v \in V$, we let $[v]$ denote the subset of $V$ containing all elements in $V$ with the same type class as $v$. The vertices $V'$ in the summary graph $G'$ are equivalence classes over the original graph  $G$, and the vertices in $V'$ are disjoint subsets of $V$. 
	$E'$ and $L'$ are, respectively, the sets of edges and labels in the graph $G'$.
	Hence, $L' \subset L$, and the elements of $E'$ are defined by the elements in the equivalence classes in $V'$ and the edges in $E$.
	Let $u, v \in V$; then $[u], [v] \in V'$.
	There is an edge $l([u], [v]) \in E'$ if and only if there exist $s \in [u] \subseteq V$ and $t \in [v] \subseteq V$ such that $l(s, t) \in E$.

	\section{Methods}\label{Methods}
	
	The Graph Summarization Problem can be considered as a problem of identifying entity type classes. The set of entity type classes can be inferred from RDF data such that each type class in the set of entity types contains the same or very similar entities only. In our method, the entity type classes are derived from the entity similarities. The discovery of the type classes, i.e., the elements $v$ in $V'$ can be also seen as clustering problem. 
	Using the calculated similarity measurements of the entity pairs, the entities that are the same or very similar, satisfying a similarity threshold, are combined in the same type class. Our entity similarity measurement approach is based on the intuition that the graph nodes that have similar relations to similar neighbors tend themselves to be similar nodes. 
	
	Previously, we investigated methods for computing entity similarities effectively. We then developed a framework for building a summary graph structure in RDF data \cite{ayvaz_building_2015,aydar_automatic_weight}. 
	This current study extends our summary graph generation approach \cite{ayvaz_building_2015} and enhances it by incorporating a measure to help discover optimum class dissimilarity thresholds and an effective method to discover the type classes automatically.
	
	\subsubsection*{Similarity of IRI Nodes}
	
	Intuitively, the characteristics of an RDF graph node are defined by its properties and the neighboring entities which are connected and related by similar properties. Based on the intuition that similar IRI nodes tend to have similar properties and interact with similar neighbor nodes, the similarities of entities in our method are calculated using the predicates of the IRI nodes, in addition to the neighbor nodes that they interact with common predicates. By neighbor we mean that a neighbor of a graph node is another node which is ``connected'' by a predicate. More formally, a node $u$ is connected to node $v$, i.e., $u$ is a neighbor of $v$, if there is a label $l \in L$ such that $l(u,v) \in E$. Therefore, a node and its neighbors are connected by a property. Thus, the similarity calculation may yield more accurate results with the addition of neighborhood similarity.  
	
	\subsubsection*{Similarity of Literal Neighbor Nodes}  
	The similarity of literal nodes indirectly impact the similarity of IRI nodes in the calculation of neighborhood similarity. The neighbors of IRI nodes can be either other IRI nodes or literals. Incorporating neighboring literals in the computation of the similarity of pairs can be beneficial, especially, in datasets where the entities are commonly described using literals. Therefore, the similarity of literal neighbor nodes are taken into account in our approach. 
	
	A literal node can consist of two or three elements: a lexical form, a datatype IRI and a language tag. The language tag in a literal node is included if and only if the datatype IRI of the literal node corresponds to rdf:langString \cite{brickley_rdf_2014}. It is important to note that the literals should be in the same language while incorporating literals in the computation of the similarity of IRI node pairs. As the same literals may have totally different meanings in different languages, we are assuming that all the literals are in the same language. If present, the rdf:langString component of the literal nodes is expected to have only one value. When calculating the similarity, the lexical form and the data type URI components a pair of literal nodes are considered. As comparing different data types is meaningless, the similarity of literal nodes is considered only when the two data types are equal.
	
	Inferring the semantics of literal nodes is challenging. To calculate the similarity of pairs of IRI nodes, an effective literal node similarity metric is needed. We make use of string similarities for the lexical form components of the literal nodes that measures common words within the two lexical forms along with their auto-generated importance weights.
	While calculating the weight of word importance in literal nodes consisting of a set of words, we consider the following factors: the source subject node, the frequency of the word within the triple collection for each subject node, and the frequency of the word within the entire dataset.
	
	\begin{equation}     
	LiteralSim( x,y )=  \frac{ \sum\limits_{i \in (x \cap y)} |t_i| \times w_i} 
	{\sum\limits_{j \in (x \cup y)} |t_j|\times w_j}
	\end{equation}   
	
	\noindent 
	where  $t$ is the term that appears in the neighbor literal nodes, such that, $u,v$ are IRI nodes in $V$, $x,y$ are literal nodes in $V$,  and  $ \enspace l(u,x), l(v,y) \in E$,  and $\sum\limits_{j \in (x \cup y)} w_j=1$. $|t_i|$ is the number of times the term appears and $w_i$ is the importancy weight of the term for the literal nodes $(u, v)$.    
	
	The importancy weight of the term for the literal nodes is calculated based on the concept of the term frequency-inverse document frequency ($tf-idf$) \cite{luhn_statistical_1957,sparck_jones_statistical_1972}, which is a widely known technique in information retrieval. $tf-idf$ indicates that some terms may be important in some documents but not as important in other documents. Said differently, the importance of a word in a document increases by its frequency in the document but its importance decreases by its frequency in the corpus \cite{rajaraman_mining_2011}.

	\subsection{Computation of Pairwise Entity Similarities}
	
	To identify the type classes in the summary graph, a metric is required to calculate the similarities of entities. For entity similarity metric, we employ an efficient graph node pair similarity metric, which utilizes the graph localities and neighborhood similarity within RoleSim similarity \cite{jin_axiomatic_2011} in conjunction with the Jaccard measure context \cite{jain_algorithms_1988}.
	
	\subsubsection*{Jaccard Similarity Measure}
	The Jaccard similarity coefficient also known as the Jaccard index is a well-known statistical measure. It is commonly used for comparing similarity and diversity of sample sets. Jaccard similarity is simply defined as the size of the intersection divided by the size of the union of the sample sets \cite{jain_algorithms_1988}.
	
	For given two sets  $S1$ and $S2$ in a dataset, the Jaccard similarity, $J(S1, S2)$, between $S1$ and $S2$ is formulated as:
	
	\begin{equation}
	J(S1, S2) = \frac{|S1 \cap S2|}{|S1 \cup S2|}
	\end{equation}
	
	When calculating the Jaccard similarity in RDF data, the subject nodes are considered to be the names or labels for the sets. Thus, the subject of the triples determine the sets. Similarly, the properties of the triples whose subject is the name or label of the set are the elements of each set. The objects of the triples whose subject is the name or label of the set become the neighbors of each subject set. 
	The object nodes, or in other words the neighboring nodes, may themselves be names or labels of sets.
	For given two subject nodes $u$ and $v$ in an RDF graph, we calculate the Jaccard similarity by noting that ${|u \cap v|}$ is the number of predicates that the subject nodes $u$ and $v$ have in common while ${|u \cup v|}$ is the number of predicates in the union of the subject nodes $u$ and $v$.

	\subsubsection*{RoleSim Similarity Measure}
	
	The Jaccard similarity has a limitation when the Jaccard index applied to an RDF graph. Because the Jaccard index determines the set similarity based on the number of common set elements only, by treating the subject nodes as sets and the predicates of the subject nodes as the set elements. However, it does not consider the relations between set elements. Thus, it does not take into account the neighboring node similarities. 
	
	For this reason, we utilize the RoleSim similarity metric which is based on the maximal matching of neighborhood pairs and a simple iterative computational method. The intuition in RoleSim similarity measure is that two nodes or entities tend to have the same role when they interact with equivalent sets of neighbors.

	Given a regular unlabeled graph $G = (V,E)$, RoleSim measures the similarity of each node pair in $V$ based on their neighborhood similarities \cite{jin_axiomatic_2011}: 
	
	\label{eqn_example} 
	
	\setlength{\arraycolsep}{0.0em}
	\begin{eqnarray}
	RoleSim(u,v) &&=(1-\beta) \\
	&&\times max_{\tiny M \in Mm(u,v)} \frac{\sum\limits_{(x,y)\in M}RoleSim(x,y)}{N_{u} + N_{v} - |M|} \nonumber\\
	&&+\beta \nonumber
	\end{eqnarray}
	\setlength{\arraycolsep}{5pt}

	$RoleSim(u,v)$ denotes the similarity of the nodes $u, v \in V$. The definition of RoleSim is recursive; i.e., $RoleSim(x,y)$ is calculated the same way as $RoleSim(u,v)$. $N(u)$ and $N(v)$ denote their respective sets of neighborhoods and $N_{u}$ and $N_{v}$ denote their respective degrees, i.e., $N_{u}$ = $|N(u)|$ and $N_{v}$ = $|N(v)|$. 
	
	$M$ is defined as a set of ordered pairs $(x,y)$ where $x \in N(u)$ and $y \in N(v)$ such that there does not exist $(x', y') \in M$, s.t. $x = x'$ or $y = y'$, and moreover, $M$ is maximal in that no more ordered pairs may be added to $M$ and keep the constraint above. $Mm(u, v)$ is the set of all such $M$'s. $Mm(u, v)$ is a set of sets.
	
	$M$ is a maximal subset of $N(u) \times N(v)$ such that no element of $N(u)$ appears more than once as a first coordinate and no element of $N(v)$ appears more than once as a second coordinate of an ordered pair in $M$. Thus, $|M| = min(N_{u},N_{v})$.
	The maximal matching ensures that the total value of selected cells has the maximum possible value. The maximal matching value, $M(u, v)$, is calculated as
	
	\begin{eqnarray}
	M(u,v) =  max_{\tiny M \in  Mm(u,v)} \frac{\sum\limits_{(x,y)\in M}RoleSim(x,y)}{Max(N_{u} , N_{v})} 
	\end{eqnarray}
	
	The parameter $ \beta$ is a decay factor, $0 < \beta < 1$. The parameter $ \beta$ is for decreasing the influence of neighbors with further distance which dampens the recursive effect.

	\subsubsection*{Combined Pairwise Entity Similarity Measure}
	
	To utilize neighborhood similarity in RDF graphs, we improve the initial Jaccard similarity by augmenting it with the RoleSim similarity measure of the neighboring nodes. When computing neighborhood similarity, comparing all neighbors to all neighbors is not an efficient method. 
	Thus, we compare only the neighboring nodes which are related by the same predicate. For instance, given two nodes $u$ and $v$, let's assume that $s_1$ and $s_2$ are neighbors of $u$, and $t$ is a neighbor of $v$. We calculate similarity of the neighborhood pairs ($s_1$, $t$) and ($s_2$, $t$) only if there is a predicate which connects $u$ to $s_1$, $u$ to $s_2$ and also connects $v$ to $t$, and we use the maximum similarity between the neighborhood pairs ($s_1$, $t$) and ($s_2$, $t$) as implied in the maximal matching concept of RoleSim similarity measure. The impact of the similarity of the neighbor nodes are weighted by each common predicate. 
	
	We note, however, that the generic version of the RoleSim measure is introduced for the unlabeled graph. In this work the input data is in RDF model, which is a directed and labeled graph. Therefore, we utilize the $RoleSim(u,v)$ measure when there
	may be multiple neighbors reached from the node pairs $u$ and $v$ by a common predicate, where $u$ and $v$ are the nodes in the input graph. 
	
	In the lists below, for $1 \leq i \leq j$, $l_{i}$ is a label for an edge, i.e., $l_{i} \in L$.  When $1 \leq h \leq j$ and if $i$ and $h$ are not equal, then $l_{i}$ and $l_{h}$ are different labels, i.e., $l_{i}$ and $l_{h}$ are different properties.  $[x_{i}]$ and $[y_{i}]$ are the sets of nodes which are related to $u$ and $v$, respectively, by predicate $l_{i}$.
	
	$l_{1}(u,[x_{1}]),l_{2}(u,[x_{2}]),...l_{j}(u,[x_{j}]) \in E$ 
	
	$l_{1}(v,[y_{1}]),l_{2}(v,[y_{2}]),...l_{j}(v,[y_{j}]) \in E$.
	
	Thus, we are assuming that there are $j$ different predicates which are predicates in triples with subject $u$ and are also predicates in triples with subject $v$.
	
	Then, by using the Jaccard index in conjunction with the RoleSim measure, their similarity can be calculated as:

	\setlength{\arraycolsep}{0.0em}
	\begin{eqnarray}
	PairSim( u,v )^{k}&& =  (1-\beta)  \\
	&& \times  \frac{1}{|u \cup v|} \nonumber  \\
	&& \times ( \sum\limits_{j \in (u \cap v)} max_{\tiny M \in Mm^{j}(u,v)} (\frac{\sum\limits_{(x,y)\in M}Sim(x,y)^{k-1} }{N^{j}_{u} + N^{j}_{v} - |M|} ) \times w_j)  \nonumber\\
	&&+\beta \nonumber
	\end{eqnarray}
	\setlength{\arraycolsep}{5pt}
	
	where $k$ is the iteration number $1 \leq k < MaxIter$, $MaxIter$ is the maximum number of iterations, such that, if $k = 3$ then $PairSim( u,v )^{k}$ denotes to the similarity of the node pair $(u, v)$ at the third iteration and $PairSim( u,v )^{k-1}$ denotes to the similarity of the node pair $(u, v)$ by the end of the second iteration. Also, $N^{j}_{(u)}$ and $N^{j}_{(v)}$ denote their respective neighborhoods that are reached by $jth$ common edge. $x \in N^{j}_{(u)}$ and $y\in N^{j}_{(v)}$, and $N^{j}_{u}$ and $N^{j}_{v}$ denote their respective degree connected by $jth$ common edge. Said differently, $N^{j}_{(u)}$ is the cardinality of $[x_{j}]$, and $N^{j}_{(v)}$ is the cardinality of $[y_{j}]$. 
	$w_j$ is the weight of the property connecting the graph nodes $(u, v)$ and their respective neighbors $(x, y)$.  
	\begin{equation}
	Sim( x, y )^{k-1}=
	\begin{cases}
	PairSim( x,y )^{k-1}, & \text{if} \enspace \text{x,y are IRI nodes } \\
	LiteralSim( x,y ), & \text{if} \enspace \text{x,y are Literal nodes } \\
	0, &  \text{otherwise }   
	\end{cases}
	\end{equation}

	We define $M$ to be a set of ordered pairs $(x,y)$ where $x \in N^{j}_{(u)}$ and $y \in N^{j}_(v)$ such that there does not exist $(x', y') \in M$, s.t. $x = x'$ or $y = y'$, and furthermore, $M$ is maximal in that no more ordered pairs may be added to $M$ and keep the constraint above. $Mm^{j}(u,v)$ is the set of all such $M$'s. $Mm^{j}(u,v)$ is a set of sets.
	
	By a \enquote{maximal nonrepeating matching}, we mean that we form as many pairs as we can from the elements in $N^{j}_{(u)}$ and  $N^{j}_{(v)}$ with the restriction that no element in either $N^{j}_{(u)}$ and  $N^{j}_{(v)}$ may be used in more than one ordered pair.
	
	The parameter $\beta$ is a decay factor $0 < \beta < 1$, which helps reduce the influence of neighbors with further distance due to the recursive effect. $l_{1}(u, x)$ and $l_{2}(v, y)$ represent directed edge labels s.t. $l_{1},l_{2}\in L,$ and $l_{1}=l_{2}$, $x \in N_{(u)}$ and  $y\in N_{(v)}$.

	\subsection{The Summary Graph Generator Algorithm}
	
	While calculating the neighborhood similarity, our proposed node similarity metric makes calls to the immediate neighbors' similarities. Since neighbors' similarities depend on their own neighbors' similarities, the immediate neighbors' similarities are not known ahead of time. 
	A solution involving recursive calls is not an efficient option in this case as it may lead to inefficient resource utilization and excessive recursion. 
	For instance, an object node $n_1$ of a subject node $n_2$ in an RDF triple may be a subject node $n_1$ of the object node $n_2$ in another RDF triple. 
	Therefore, our algorithm runs in multiple iterations until the rate of change in calculated similarities drops under a given threshold. It is a similar approach to the PageRank algorithm \cite{page_pagerank_1999}. The initial similarity of a node pair is set to 1 if they share a common predicate and 0, otherwise. Such that:  
	\\*
	$\forall(u,v \in V) :$
	\\* 
	$(S(u,v) = 1) \rightarrow (|u \cap v| > 0)$ and,
	\\*
	$(S(u,v) = 0) \rightarrow (|u \cap v| = 0) $ 
	\\*\\*
	
	Our approach is two folds. Firstly, the pairwise similarity algorithm calculates the similarity values for each pair which constructs a similarity matrix. Once the pairwise similarities converge, the type class generation algorithm begins and generates the type classes i,e, it assigns the node $u$ and $v$ to the same type class if their dissimilarity value is less than an auto calculated 
	$\epsilon$ threshold which is the class dissimilarity threshold.  
	
	As the algorithm generates common pairs if two candidate nodes that share at least one common predicate, the overall complexity for the algorithm is $n^2$ in the worst case. It occurs when all subject nodes in the RDF graph have a common predicate with every other subject node.
	When the noise predicates excluded, i.e. the predicates that is referenced by most if not all the subject nodes, the algorithm performs better than $n^2$.
	Thus, the complexity of the algorithm depends on the characteristics of the dataset. On a dense graph, the complexity approaches to $n^2$ while it gets near to $n(logn) k$ time in sparse graphs, where $k$ is a constant number of iteration.

	The basic steps of the algorithm include sorting the triples according to their predicate label,  extraction of the subject node pairs for each of the predicates,  running the similarity computation algorithm in iterations until convergence and generating the type classes based on the calculated similarity measures.
	
	\begin{algorithm}
		\SetKwInOut{Input}{input}	
		\SetKwInOut{Output}{output}
		\SetKwInOut{Parameter}{parameter}
		\Input{Graph $G(V,L,E)$ }
		\Output{Similarity-Matrix $S$, Pairs $H$}
		\Parameter{MaximumIteration $MaxIter$, Iteration-Convergence-Threshold  $Ict$}
		\BlankLine	
		 $\forall pair(u,v)  \in V \space\space (S(u,v) \gets 0 )$;\BlankLine
		 $H \gets \emptyset $;\BlankLine
		 $T \gets Sort(E) $ by $l,u,v$ s.t. $ l(u,v) \in E $;\BlankLine	
		
		\For{$\textbf{each} $ distinct $ pair(u,v)$ from $ T $}{
			\If{$ \exists( l_{1}(u,x) $ and $ l_{2}(v,y)  ) $ s.t. $l_{1}=l_{2}$ }{
				 $S(u,v) \gets 1$\BlankLine	
				 $P(u,v) \gets (u,v,L^{j},N^{j}(u),N^{j}(v))$ where $u,v \in V$,  $L^{j}$ is the list of common labels between $u$ and $v$, and $L^{j} \in L$ \BlankLine	
				 $H \gets H \cup \{P(u,v)\} $\BlankLine	
			}
		}
		 $S_{previous} \gets \emptyset$\BlankLine	
		 $converged \gets false$\BlankLine	
		 $count \gets 0$\BlankLine	
		\While {$converged = false $ and $count < MaxIter $}{
			\For{$\textbf{each} ( (u,v,L^{j},N^{j}(u),N^{j}(v)) \in H )$}{
				 ${ PairSim( u,v )^{k} =  (1-\beta)  
					\times  \frac{1}{|u \cup v|} \nonumber  
					\times ( \sum\limits_{j \in (u \cap v)} max_{\tiny M \in Mm^{j}(u,v)} (\frac{\sum\limits_{(x,y)\in M}Sim(x,y)^{k-1} }{N^{j}_{u} + N^{j}_{v} - |M|} ) \times w_j)   
					+\beta \nonumber} $\BlankLine	
				 $S(u,v) \gets PairSim( u,v )^{k}$	
			}
			 $converged = |S - S_{previous}| \leq Ict$ \BlankLine
			 $S_{previous} \gets S$ \BlankLine
			 $count \gets count+1$ \BlankLine
		} 
		\Return $S,H$
		
		\caption{SimMeasure}          
		\label{Schema Graph with Jaccard&Rolesim}        
	\end{algorithm}
	
	The type class generation algorithm creates distinct classes, such that subject node pairs that have similarity greater than a given threshold get put to the same type class. The input parameter $\beta$ is a decay factor $0 < \beta < 1$. $l_{1}(u, x)$ and $l_{2}(v, y)$ represent directed edge labels s.t. $l_{1},l_{2}\in L,$ and $l_{1}=l_{2}$, $x \in N_{(u)}$ and  $y\in N_{(v)}$.

	\begin{algorithm}               
		\SetKwInOut{Input}{input}	
		\SetKwInOut{Output}{output}
		\SetKwInOut{Parameter}{parameter}
		\Input{Similarity-Matrix $S$, Pairs $H$}
		\Output{Auto-Generated-Type-Classes-Map $C$ }
		\Parameter{Class-Dissimilarity-Threshold $\epsilon$}
		\BlankLine	
		
		\caption{CreateClasses}          
		\label{Class creation algorithm}        
		
		\For{$\textbf{each} ( (u,v,L^{j},N^{j}(u),N^{j}(v)) \in H )$}{
			\If {$C(u)$ exists }{
				 $ c_{i} \gets C(u) $\BlankLine
				\Else{
					 $ c_{i} \gets \{u\} $
				}
			}
			\If {$1-S(u,v) < \epsilon $ }{
				\If {$C(v)$ exists }{
					 $ c_{i} \gets c_{i} \cup C(v) $
					\Else{
						 $ c_{i} \gets c_{i} \cup \{v\} $
					}
				}
				 $ c_{i} \gets C(v) $\BlankLine
				
				 $ C(u) \gets c_{i} $\BlankLine
				 $ C(v) \gets c_{i} $\BlankLine
			}
		} 
		\Return $C$
	\end{algorithm}

	\subsection{Dynamic Assignment of Weights of IRI Node Descriptors }
	An IRI node is described through its predicates and the collection of literal neighboring nodes in the lexical form. For simplicity, we call the predicates and literal neighboring nodes as descriptors of the IRI nodes. As stated above, the similarity of a pair of IRI nodes depend upon their descriptor similarities and the similarities of their neighbors. 
	
	The weight of each descriptor may vary significantly as each descriptor may have different impact on an IRI node.
	Hence, identifying appropriate metrics for generating IRI descriptor weights is a vital task in computation of accurate
	similarity values.
	
	Upon investigations on the factors that can impact the weight of a descriptor, we propose an approach in this study for generating the importance weights of the IRI node descriptors automatically. Based on the investigations, we think that the weight of a descriptor may differ for each IRI for which it is a descriptor and the weight increases proportionally by the number of times a descriptor appears in the reference IRI, but it is offset by the frequency of the descriptor in the entire RDF dataset. 
	This tendency is similar to the concept of the term frequency-inverse document frequency ($tf-idf$).
	While computing the weight of properties dynamically, we apply the $tfidf$ to the properties and nodes in RDF graphs. $tfidf$ is calculated as follows:
	
	\begin{equation} 
	tfidf(p,u,G) = tf(p,u) \times idf(p, G).
	\end{equation}
	
	\noindent
	where the term frequency (tf) \cite{luhn_statistical_1957} represents the frequency of a proposition
	$p$ with respect to a graph subject node $u$. More exactly, when $u \in V$
	and $p \in L$, then 
	
	\begin{equation}  
	f(p, u) = | \{ v \in V : p(u, v) \in E \} |.
	\end{equation}
	
	\noindent
	
	Equivalently, $f(p, u)$ is the number of RDF triples with subject $u$ and
	property $p$.
	
	\smallskip\
	
	To define $tf(p, u)$, it is helpful to have a notation for the set of all
	properties with subject $u$.  Thus, for $u \in V$, 
	$ L(u) = \{ q \in L : \exists v \in V \textrm{ with } q(u, v) \in E \}$.
	Then

	\begin{equation}  
	tf(p,u) =  \frac{ f(p, u)}{ \sum_{q \in L(u)} f(q, u) }.  
	\end{equation}

	The inverse document frequency (idf) \cite{sparck_jones_statistical_1972} represents the frequency of a property usage across all graph nodes, and it is defined as
	
	\begin{equation} 
	idf(p, G) =  \ln \frac{|V|}{|\{u \in V: p \in L(u)\}|}. 
	\end{equation}

	The property importance weights are based on the degree of distinctiveness of a property describing an entity. With property distinctiveness, we mean the uniqueness of a property in describing the key characteristics of an entity type.
	For instance, if a property is specific to an entity type, it is a distinguishing character of the type from other types. When a property exists in all entity types, its quality of being distinctive is low. The noise labels tend to be common for a majority of entities if not for all entities. By increasing importance weights of properties with a higher degree of distinctiveness, we reduce the importance of noise labels automatically. As a result, the noise labels have significantly less impact on the overall similarity measures.

	\subsection{Class Predicate Stability} 
	
	In this work, the summary graph is built automatically from an RDF dataset. However, automatically generated summary graphs can be error prone. It is essential to have an effective metric to measure the degree of confidence of a relation between classes in the summary graph. We define this metric as Class Predicate Stability (CPS), which is a similar notion to the concept of stability that introduced by Paige and Tarjan \cite{paige1987three}. 
	
	For $u$ and $v$ being IRIs in the dataset, $G=(V,E,L)$, and $u \in c_1$ and $v \in c_2$ with both $c_1$ and $c_2$ being type classes in the summary graph, $ G'=(V',E',L')$, a class relation between the class $c_1$ and the class $c_2$ is generated as a predicate and represented as $l(c_1,c_2)$ if there is at least one relation $l(u,v)$. Consequently, $l \in L'$ and $l(c_1, c_2) \in E'$. 
	
	The CPS metric is calculated as the number of the IRI nodes $u$ in class $c_1$ having a triple of the form $(u, p, v)$ with $u \in c_1$ and $v \in c_2$ divided by the total number of the IRI nodes in $c_1$ in the summary graph such that the triple $(c_1, p, c_2)$ is in the summary graph $G'$ and $c_1$ and $c_2$ are type class IRI nodes with $p$ being a predicate between them. $CPS(c_1,p,c_2)$ is formulated as
	
	\begin{equation}
	CPS(c_1,p,c_2) = \frac{|(u,p,v):u \in c_1, v \in c_2\}|}{ |c_1|}   
	\end{equation}
	where $|c_1|$ is the number of IRI nodes in the class $c1$. Note that $|c_1| > 0$.
	
	The CPS value for a triple $(c_1, p, c_2)$ indicates the degree of partitioning coarseness of the type classes $c_1$ and $c_2$ with the predicate $p$ in the summary graph. Hence, the mean of all the CPS values in the summary graph is an indicator of accuracy for the generated summary graph. $CPS(G')$ is formulated as

	\begin{equation}
	CPS(G') = \frac{\sum\limits_{i=1}^{|E'|} CPS(c_1^i,p^i,c_2^i)}{|E'|}     
	\end{equation}
	where $ G'=(V',E',L')$ is the summary graph and $ p^i(c_1^i,c_2^i) \in E' $, and thus $|E'| > 0$.   
	\newline
	For two classes $c_1$ and $c_2$ in the summary graph, when either all the IRI nodes from $c_1$ are connected with a predicate $p$ to at least one IRI node in $c_2$ or none of the IRI nodes in $c_1$ are connected with the predicate $p$ to an IRI node in $c_2$, we call that the classes $c_1$ and $c_2$ have full CPS.

	\subsection{Automatic Calculation of the Class Dissimilarity Thresholds}\label{Automatic Calculation of the Class Dissimilarity Thresholds} 
	
	Our approach automatically builds the summary graph from RDF data. A drawback in the automatic summary graph generation approach is the need for estimating the optimum parameters that help determine the type classes. As expected, higher class dissimilarity threshold generates more coarse classes, whereas the classes become more granular when the threshold is chosen smaller. 
	The optimum values for  the class dissimilarity threshold depend on the characteristics of the datasets. In real-world datasets, users may not have a good grasp on the underlying data to determine optimal class dissimilarity threshold values.   
	
	To determine how closely the entities fit the type class, an effective metric is needed to measure the degree of fit within each type class in the summary graph. For this purpose, we utilize the root-mean-square deviation (RMSD), which is a commonly used measure of the differences between the values in comparison \cite{levinson1946wiener}. 
	
	\subsubsection*{The root-mean-square deviation (RMSD) in RDF summary graphs}
	
	The RMSD represents the amount of the deviations of IRI node property values from the class center and provides a single measure of predictive power. In RDF summary graph, we calculate the overall RMSD by aggregating the sum of RMSD values for each type class in the summary graph.   
	
	To calculate the RMSD of summary graph, we first determine the centroids for each type class and then compute the RMSD between the class centroids and all IRI nodes within the type class using Manhattan distance. In RMSD calculation, the IRI node properties represent the dimensions of the IRI nodes within the type class. $RMSD(G')$ of summary graph $G'$ is formulated as follows
	
	\begin{equation}
	RMSD(G') = \sum\limits_{c_i \in G'}{ \sqrt{\frac{\sum\limits_{(i=1) \in L'}^{n} (x_i - \bar{x}) }{n} } }
	\end{equation}
	where $c_i$, $L'$ are, respectively, the list of classes and the property labels in the summary graph $G'$. $x_i$ represents the IRI nodes in type class and $\bar{x}$ denotes the centroid for members of a particular type class in the summary graph $G'$.

	Higher RMSD values in a summary graph indicate that entities within type classes sparsely located. When the entities in type classes are very similar to each other, the center of the cluster will be dense. Thus, the sum of distances to the centroids and the cumulative RMSD value will be lower accordingly. 
	
	\subsubsection*{Discovery of Class Dissimilarity Threshold}
	
	To discover the type classes in summary graph automatically, we propose a measure, called Favorability, to calculate the class dissimilarity threshold automatically as follows. 
	
	\begin{equation}
	Favorability(G') = max\left\lbrace { \frac{ Stability(G') * Typification Rate(G') }{ (RMSD(G') + 0.1)} }\right\rbrace 
	\end{equation}

	\begin{algorithm}
		\SetKwInOut{Input}{input}	
		\SetKwInOut{Output}{output}
		\SetKwInOut{Parameter}{parameter}
		\Input{Similarity-Matrix $S$, Pairs $H$, 
			Minimum-Threshold $min_{\tiny \epsilon}$, Maximum-Threshold $max_{\tiny \epsilon}$, Number-of-try $n$,
			Previous-Favorability $prev\_favor$, Previous-Optimum-Threshold $prev\_optimum_{\tiny \epsilon}$}
		\Output{Optimum-Threshold $optimum_{\tiny \epsilon}$}
		\Parameter{Epsilon-Convergence-Threshold $Ect$}
		\BlankLine	
		\caption{$FindOptimumEpsilon$}
		\label{Finding optimum class dissimilarity threshold algorithm}        
		
		 $current_{\tiny \epsilon} \gets min_{\tiny \epsilon}$\BlankLine	
		
		 $inc \gets (max_{\tiny \epsilon} - min_{\tiny \epsilon}) / n$\BlankLine	
		
		 $optimum_{\tiny favor} \gets prev\_favor$\BlankLine	
		 $optimum_{\tiny \epsilon} \gets prev\_optimum_{\tiny \epsilon}$\BlankLine	
		
		\While {$current_{\tiny \epsilon} \leq max_{\tiny \epsilon} $}{
			
			 $(G', C) \gets CreateClasses(S, H, current_{\tiny \epsilon})$\BlankLine 
			
			 $Favorability(G') =  \frac{ Stability(G') * Typification Rate(G') }{ (RMSD(G') + 0.1)} $\BlankLine
			
			\If {$Favorability(G') > optimum_{\tiny favor} $}{
				 $optimum_{\tiny favor} \gets Favorability(G')$\BlankLine
				 $optimum_{\tiny \epsilon} \gets current_{\tiny \epsilon}$
			}		
			 $current_{\tiny \epsilon} \gets current_{\tiny \epsilon} + inc $
		}	
		\If {$ |(optimum_{\tiny favor} - prev\_favor)| > Ect $}{
			 $optimum_{\tiny \epsilon} \gets FindOptimumEpsilon(S, H, optimum_{\tiny \epsilon} - inc , optimum_{\tiny \epsilon} + inc, n/2, optimum_{\tiny favor}, optimum_{\tiny \epsilon} ) $		      
		}	
		\Return $optimum_{\tiny \epsilon}$
		
	\end{algorithm}
	
	The idea behind the formula is that we think that the quality of summary graph type classes is associated and directly proportional to the summary graph stability, a measure of relationships between type classes, and the ratio of entities belonging to a type class, and inversely proportional to the RMSD, the degree of inner class deviation.
	
	In the formula, $Favorability(G')$ is the class dissimilarity threshold for the summary graph $G'$ and $Typification Rate(G')$ represents the rate of entities that belong to a type class based on the class dissimilarity threshold. The $Typification Rate(G')$ is low when the class dissimilarity threshold is too high since the number of entities satisfying high similarity threshold for class membership will be small. 
	
	To obtain the high quality results while reducing the computation cost, we gradually change the threshold values in constant number of times and set the class threshold value that provides the maximum value of the proposed measure. The algorithm \ref{Finding optimum class dissimilarity threshold algorithm} demonstrates how the optimum class dissimilarity threshold is discovered utilizing the favorability measure. In the algorithm, $optimum_{\tiny \epsilon}$ refers to the optimum class dissimilarity threshold for the summary graph.
	
	The proposed automatic threshold discovery measure is not assumed to be perfect. Finding optimum summary graph type classes is a formidable problem as the quality of summary graph is dependent on the type of datasets.  Despite this, the proposed measure integrates different aspects of the graph summaries and provides intuitively accurate graph summaries based on our evaluations.

	\section{Evaluations}\label{Evaluations}
	In the evaluations, we conducted preliminary experiments on three datasets: a subset of DBpedia \cite{auer_dbpedia:_2007}; a subset of SemanticDB \cite{d2012semanticdb}, and a subset of Lehigh University Benchmark (LUBM) \cite{guo2005lubm}. Our experimental datasets are in different domains and they represent different aspects of real world semantic data.
	
	SemanticDB is a Semantic Web content repository for Clinical Research and Quality Reporting in cardiovascular surgery domain. The structured entity type information exist in SemanticDB which we utilized as the ground truth for the verification of the algorithm. Lehigh University Benchmark (LUBM) is a well-known benchmark for OWL knowledge base systems, which also has entity type information available. But unlike SemanticDB, LUBM data has hierarchical types. Lastly, DBPedia a central source in the Linked Open Data Cloud \cite{bizer_linked_2009} and is a commonly used general purpose dataset. However, using the entity type information for the verification of the algorithm is more problematic in DBPedia, as the type information may not present, or an entity may have several types including the hierarchical types. Therefore, we manually verified the results of the algorithm. Table \ref{tab:triples} demonstrates a sample of RDF triples from each dataset in the evaluations.

	\begin{table} 
		\footnotesize 
		\caption{A Sample of RDF Triples from Each Dataset} 
		\scalebox{0.77}	
		{%
			\begin{tabular}{p{2cm} p{3.3cm} l l} 
				\hline\hline 
				\\[0.2ex]
				Dataset  &  Subject  & Predicate & Object \\[0.5ex]
				\hline 
				\\[0.01ex]
				SemanticDB & SurgeryProcedure:236 & SurgeryProcedureClass & "cardiac valve" \\[1ex]
				SemanticDB & SurgeryProcedure:236 & CardiacValveEtiology & "other" \\[1ex]
				SemanticDB & SurgeryProcedure:236 & belongsToEvent & Event:184 \\[1ex]
				SemanticDB & SurgeryProcedure:236 & SurgeryProcedureDescription & "pulmonary valve repair" \\[1ex]
				SemanticDB & SurgeryProcedure:236 & CardiacValveStatusologyData & "native" \\[1ex]
				SemanticDB & SurgeryProcedure:104 & SurgeryProcedureClass & "cardiac valve" \\[1ex]
				SemanticDB & SurgeryProcedure:104 & CardiacValveEtiology & "rheumatic" \\[1ex]
				SemanticDB & SurgeryProcedure:104 & belongsToEvent & Event:81 \\[1ex]
				SemanticDB & SurgeryProcedure:104 & SurgeryProcedureDescription & "mitral valve repair" \\[1ex]
				SemanticDB & SurgeryProcedure:104 & CardiacValveStatus & "native" \\[1ex]
				LUBM & Student49 & telephone & "xxx-xxx-xxxx" \\[1ex]
				LUBM & Student49 & memberOf & http://www.Department3.University0.edu \\[1ex]
				LUBM & Student49 & takesCourse & Course32 \\[1ex]
				LUBM & Student49 & name & "UndergraduateStudent49" \\[1ex]
				LUBM & Student49 & emailAddress & "Student49@Department3.University0.edu" \\[1ex]
				LUBM & Student49 & type & UndergraduateStudent \\[1ex] 
				LUBM & Student10 & telephone & "xxx-xxx-xxxx" \\[1ex]
				LUBM & Student10 & memberOf & http://www.Department3.University0.edu \\[1ex]
				LUBM & Student10 & takesCourse & Course20 \\[1ex]
				LUBM & Student10 & name & "UndergraduateStudent10" \\[1ex]
				LUBM & Student10 & emailAddress & "Student10@Department3.University0.edu" \\[1ex]
				LUBM & Student10 & type & UndergraduateStudent \\[1ex] 
				DBPedia & Allen\_Ginsberg & wikiPageUsesTemplate & Template:Infobox\_writer\\[1ex] 
				DBPedia & Allen\_Ginsberg & influenced & John\_Lennon\\[1ex]  
				DBPedia & Allen\_Ginsberg & influences & Fyodor\_Dostoyevsky\\[1ex]  
				DBPedia & Allen\_Ginsberg & deathPlace & "New York City, United States"@en\\[1ex] 
				DBPedia & Allen\_Ginsberg & deathDate & "1997-04-05"\\[1ex]  
				DBPedia & Allen\_Ginsberg & birthPlace & "Newark, New Jersey, United States"@en\\[1ex]  
				DBPedia & Allen\_Ginsberg & birthDate & "1926-06-03"\\[1ex]   
				DBPedia & Allen\_Ginsberg & deathPlace & "New York City, United States"@en\\[1ex] 
				DBPedia & Albert\_Camus & wikiPageUsesTemplate & Template:Infobox\_philosopher\\[1ex] 
				DBPedia & Albert\_Camus & influenced & Orhan\_Pamuk\\[1ex]  
				DBPedia & Albert\_Camus & influences & Friedrich\_Nietzsche\\[1ex] 
				DBPedia & Albert\_Camus & deathPlace & "Villeblevin, Yonne, Burgundy, France"@en\\[1ex] 
				DBPedia & Albert\_Camus & deathDate & "1960-01-04" \\[1ex] 
				DBPedia & Albert\_Camus & birthPlace & "Drean, El Taref, Algeria"@en\\[1ex] 
				DBPedia & Albert\_Camus & birthDate & "1913-11-07"\\[1ex] 
				
				\hline 
			\end{tabular}
		}
		\label{tab:triples} 
	\end{table}

	\subsection{Assessing Algorithm Parameters}
	
	We tested several parameters of the algorithm, including the maximum iteration, beta factor, class dissimilarity threshold, iteration convergence threshold ($Ict$), and the size of dataset in type generation. The results of our evaluations are demonstrated in Table \ref{tab:evaluationparams}. For verification, we extracted the ground truth, entity types present in the datasets, against the entity type classes generated by the algorithm. For the assessment of our evaluations, we used the measure of precision. Precision is defined as the ratio of correct results over all results. 
	
	\begin{table}[ht]
		\caption{Evaluations of Algorithm Parameters} 
		{
			\begin{tabular}{p{1.9cm} r c c r r } 
				\hline\hline 
				\\[0.2ex]
				Dataset & $\#$Triples & Class\_Threshold & $\#$Iterations & Stability & Precision  \\[0.5ex]
				\hline  
				\\[0.01ex]
				SemanticDB & 6,450 & 0.5 & 4 & 61.0\% & 87.3\%\\[1ex]
				LUBM & 6,484 & 0.3 & 3 & 67.8\% & 90.7\%\\[1ex]
				DBPedia & 10,000 & 0.6 & 3 & 82.4\% & 92.8\%\\[1ex]
				\hline 
			\end{tabular}
			\label{tab:evaluationparams}
		}
		
	\end{table}
	
	The similarity computation algorithm stops the iterations, once the rate of change in the similarity measures drops below the threshold or once it reaches the maximum number of iterations. In our evaluations, we observed that the similarity measures typically converge after a few iterations with the values of the maximum number of iterations and the iteration convergence threshold being as 10 and 0.001, respectively.
	
	\subsection{Performance of dynamic assignment of descriptor weights}
	
	We also evaluated the performance of dynamic assignment of descriptor weights. Table \ref{tab:propertyweights} shows a sample of dynamically assigned descriptor weights from each dataset. As expected, the algorithm assigned higher weights to the properties with a higher degree of distinctiveness describing the resource type. For instance in LUBM dataset, takesCourse property is more descriptive of the Student type than the name property, which is a common property for all type classes in the dataset. Thus, takesCourse was assigned a weight of 44.1\% as compared to the weight of 7.5\% for name.
	
	\begin{table}[ht]
		\caption{An Excerpt from Dynamically Assigned Weights of Descriptors} 
		\scalebox{0.75}
		{%
			\begin{tabular}{p{1.9cm} p{4.6cm} p{2.5cm} p{3.2cm} c } 
				\hline\hline 
				\\[0.2ex]
				Dataset & Node\_Pair  & Descriptor\_Type & Descriptor & Weight \\[0.5ex]
				\hline 
				\\[0.01ex]
				LUBM & (Student49,Student10) & Property & memberOf & 14.7\%\\[1ex]
				LUBM & (Student49,Student10) & Property & takesCourse & 44.1\%\\[1ex]
				LUBM & (Student49,Student10) & Property & emailAddress & 14.0\%\\[1ex]
				LUBM & (Student49,Student10) & Property & type & 5.7\%\\[1ex]
				LUBM & (Student49,Student10) & Property & name & 7.5\%\\[1ex]
				LUBM & (Student49,Student10) & Property & telephone & 14.0\%\\[1ex]
				SemanticDB & (Procedure:236,Procedure:104) & Literal & "cardiac" & 13.6\%\\[1ex]
				SemanticDB & (Procedure:236,Procedure:104) & Literal & "native" & 15.2\%\\[1ex] 
				SemanticDB & (Procedure:236,Procedure:104) & Literal & "other" & 14.3\%\\[1ex] 
				SemanticDB & (Procedure:236,Procedure:104) & Literal & "pulmonary" & 22.8\%\\[1ex] 
				SemanticDB & (Procedure:236,Procedure:104) & Literal & "repair" & 17.2\%\\[1ex] 
				SemanticDB & (Procedure:236,Procedure:104) & Literal & "valve" & 16.9\%\\[1ex]  
				DBPedia & (Allen\_Ginsberg,Albert\_Camus) & Property & wikiPageUsesTemplate & 2.2\%\\[1ex]  
				DBPedia & (Allen\_Ginsberg,Albert\_Camus) & Property & influences & 58.3\%\\[1ex] 
				DBPedia & (Allen\_Ginsberg,Albert\_Camus) & Property & deathDate & 2.2\%\\[1ex] 
				DBPedia & (Allen\_Ginsberg,Albert\_Camus) & Property & birthDate & 2.4\%\\[1ex] 
				DBPedia & (Allen\_Ginsberg,Albert\_Camus) & Property & birthPlace & 2.1\%\\[1ex] 
				DBPedia & (Allen\_Ginsberg,Albert\_Camus) & Property & deathPlace & 2.1\%\\[1ex] 
				DBPedia & (Allen\_Ginsberg,Albert\_Camus) & Property & influenced & 30.7\%\\[1ex] 
				
				\hline 
			\end{tabular}
		}
		\label{tab:propertyweights} 
	\end{table}

	\subsection{Effectiveness of the Automatic Computation of Class Thresholds}
	
	In a set of evaluations, we further assessed the effectiveness of automatic calculation of the class threshold approach using a subset of the same set of datasets. As demonstrated in Table \ref{tab:evaluationauto}, the stability, RMSD and optimum class threshold may vary depending on the characteristics of the datasets. In LUBM dataset, the RMSD result was higher compared to the other datasets. Among them, the highest optimum class dissimilarity threshold, 0.56, was achieved in DBPedia dataset. This means that the entities within the type classes of the summary graph generated by the dataset contained similar properties and were very similar to the entities that belonged to the same type class. In our evaluations, we observed that the epsilon convergence threshold around 0.9 performed well.

	\begin{table}[ht]
		\caption{Automatic Calculation of the Class Thresholds Results} 
		{
			\begin{tabular}{p{1.9cm} r c c r r } 			
				\hline\hline 
				\\[0.2ex]
				Dataset & $\#$Triples & Optimum\_Class\_Threshold & RMSD & Stability \\[0.5ex]
				\hline  
				\\[0.01ex]
				SemanticDB & 1000 & 0.32 & 0.8 & 74.0\% \\[1ex]
				LUBM & 1000 & 0.26 & 4.8 & 83.0\% \\[1ex]
				DBPedia & 1000 & 0.56 & 0.9 & 82.8\% \\[1ex]
				\hline 
			\end{tabular}
			\label{tab:evaluationauto}
		}
	\end{table}
	
	\subsection{Generated Summary Graph}
	
\begin{figure} [ht]
		\centering
		\includegraphics[height=8.5 cm]{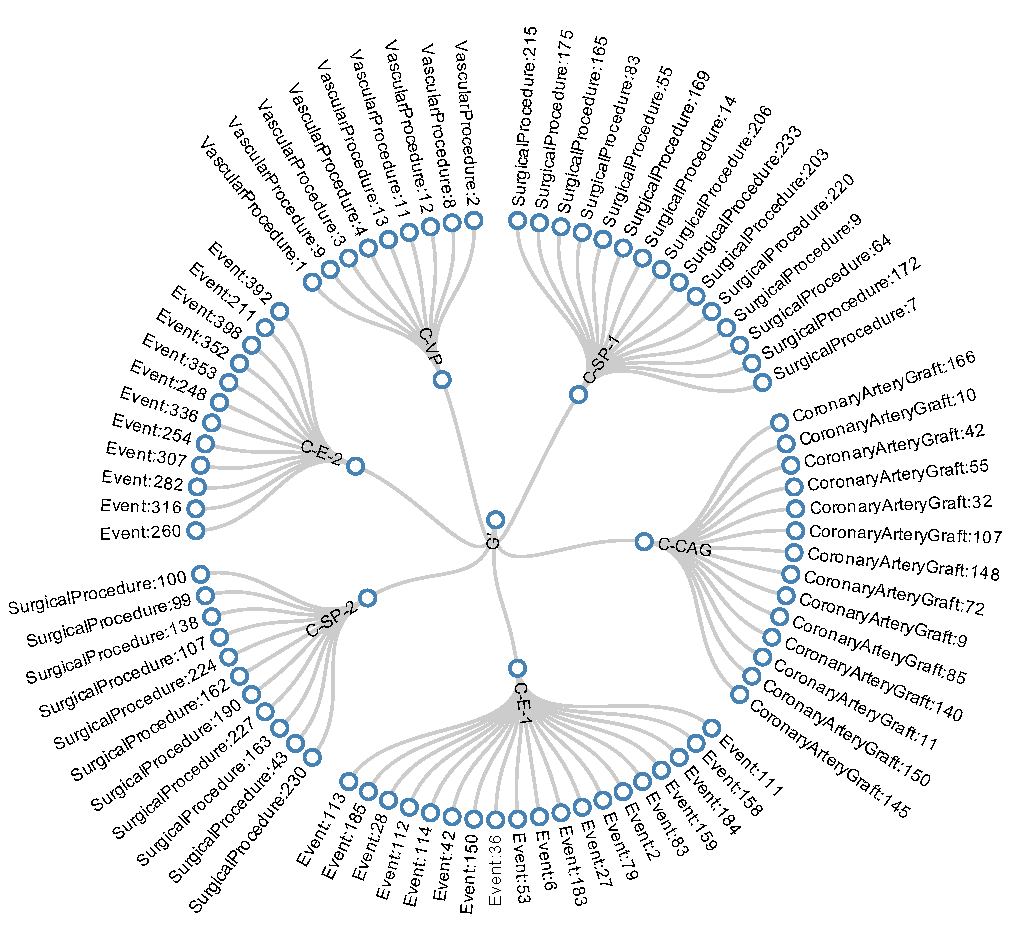}
		\caption{A figure consisting of different types of
			entities and elements belonging to the type classes.}
		\label{fig:figure_categories}
	\end{figure}

	Figure \ref{fig:figure_categories} illustrates a small sample set of entities in the RDF graph from SemanticDB and their corresponding type classes in the summary graph. As demonstrated in Figure \ref{fig:figure_categories}, the classes C-E1 and C-E2 represent the entities that are patient event types. They are classified in two different classes because when compared with the original dataset we observed that the entities in C-E1 are more specifically patient surgery-related event types while the entities in C-E2 are patient-encounter related event types. Also, the classes E-SP1 and E-SP2 are surgical procedure types. More specifically, the entities in E-SP1 are coronary artery and vascular procedure-related procedures while the entities in E-SP2 are cardiac valve related-procedures. The classes C-VP and C-CAG represent the entities that are related to vascular procedures and coronary artery grafts, respectively. We implemented a basic algorithm to name the classes based on the class member IRIs. The classes C-E1, C-E2, C-SP1, C-SP2, C-VP and C-CAG are named as C-Event-1, C-Event-2, C-SurgicalProcedure-1, C-SurgicalProcedure-2, C-VascularProcedure and C-CoronaryArteryGraft, respectively.
	
\begin{figure}
		\centering
		\includegraphics[height=8.5 cm]{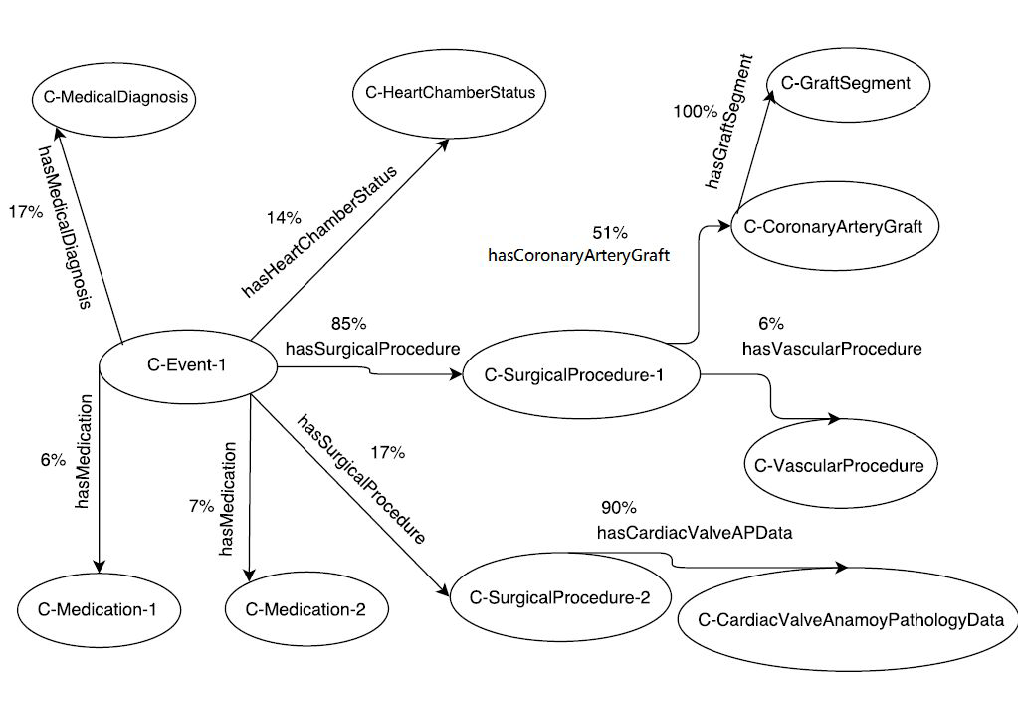}
		\caption{An excerpt from the generated summary graph.}
		\label{fig:excerpt_sdb_summarygraph}
	\end{figure}

	The summary graph is generated along with the classes and the class relations with a stability measure for each relation. Figure \ref{fig:excerpt_sdb_summarygraph} shows an excerpt from the summary graph representing the class relations from SemanticDB dataset. The percentage values beside the predicates are the stability (CPS) measure.

	Overall, we observed that the class dissimilarity threshold ranging between 0.25 to 0.6 with the beta factor of 0.15 appeared to work well in our evaluations. The automatically calculated class dissimilarity threshold values during the evaluations were in close proximity of the threshold values for the datasets that were kept as the ground truth in the assessment.  
	
	\section{Limitations of the method}
	
	The algorithm used for the similarity calculation runs in the $n^2$ in the worst case and in the $n(logn) k$ time in average, where $k$ is a constant number of iterations. For Web-scale usage, the scalability of the algorithm needs to be further improved as the size of the input RDF data can be very large. For instance, as of today, the Linking Open Data\cite{bizer_linked_2009} project already contains more than 30 billions triples. In future work, we plan to address the performance issues for big datasets in the worst-case scenario and perform Web-scale evaluations.
	
	Furthermore, the literal node similarity calculation currently does not perform well in cases where the literal nodes belong to different languages with disparate linguistic properties as we do not perform any linguistic analysis.
	
	Also, in the current study, the classes in the summary graph are automatically named exploiting the frequent entity names and literal values that belong to the related class. The naming method may not always generate the best names for human readers.
	
	\section{Related Work}\label{Related Work}
	
	The problem of Graph Summarization has been studied by various communities from different perspectives including Graph compression, graph partitioning, social network analysis, data visualization.
	
	From the Graph Compression perspective, numerous approaches have explored the graph summarization problem with the aim of reducing the storage space of the large graph datasets \cite{raghavan2003representing,fan2012query,he2000fast,chierichetti2009compressing}. Different from these approaches, we deal with labeled directed graphs as in the case of RDF. Also, a summary graph structure based on the original graph is generated in our method. 
	
	Several studies such as \cite{newman2003structure,chakrabarti2006graph} have broadly investigated statistical methods to help understand the properties of large networks. These approaches provide useful information but they do not generate a summary graph from the graph data as it is the focus of our approach. 
	
	In the area of graph partitioning area, many methods have been introduced \cite{newman2004finding,xu2007scan,tian2008efficient,zhang2010discovery} to partition graph data into specific components. While these methods are helpful in discovering neighborhoods in large graph networks, they don't consider the similarities of the node properties. 
	Tian et al. \cite{tian2008efficient,zhang2010discovery} proposed an aggregation-based graph summarization utilizing graph node attributes. However, the approach only deals with categorical node attributes and users need to group numerical attributes into categories manually, which is not feasible for large real-world datasets.   
	
	In the Semantic Web community, there has also been some related studies \cite{campinas2012introducing,khatchadourian2010explod,tran2010structure,ayvaz_building_2015,goasdoue2015query}. The studies in \cite{campinas2012introducing} and \cite{goasdoue2015query} are query driven graph summarization methods. They primarily focus on the problem of SPARQL query formulation over RDF data. The approaches using bisimulation \cite{khatchadourian2010explod,tran2010structure} have a  limitation to be applied in real-world datasets due to the exponential complexity of bisimulation. 
	
	Neighborhood-based similarity measures have been investigated by several studies including SimRank \cite{jeh_simrank:_2002}, SimRank++ \cite{antonellis2008simrank++}, PageSim \cite{lin2006pagesim}, MatchSim \cite{lin_matchsim:_2009}, PathSim \cite{sun2011pathsim}, and  Co-Citation \cite{small1973co}. Especially, SimRank is a widely-known measure, which utilizes the mean of the edge similarities between nodes. However, this may reduce the similarity score of similar graph nodes in a counterintuitive manner when the nodes have multiple edges that differ in weights. On contrary, our method considers the maximal matching for calculating the similarity in a structural context.
	
	Entity properties might have different impact on entity similarity scores. The weights of the entity properties can be determined using a similarity measure.  There are some studies that try to calculate the property weights and apply them in similarity calculations such as \cite{seddiqui_efficient_2015,castano_instance_2008}. But, they primarily focus on instance matching. In instance matching, the property weights yield precedence to properties making the instances more unique. Contrary to instance matching, the properties that would help describe the entity types more distinctively are weighted higher in our approach. In \cite{castano_instance_2008}, they determine the property weights using the distinct value based weight generation and assign higher weight to a property that references more distinct values. However, a training set of instances may not always be available.

	\section{Conclusion}\label{Conclusion}
	In this paper, we have investigated the main aspects for graph summary problem in RDF graphs. 
	We described our pairwise graph node similarity calculation with the addition of the property and string word importance weights, along with the Class Predicate Stability metric, which allows evaluation of the degree of confidence of each class predicate in the summary graph. 
	Furthermore, we studied obtaining the optimum value of the class dissimilarity threshold automatically in RDF summary graphs.
	Based on our investigations, a measure to determine optimum class dissimilarity thresholds and an effective method to discover the type classes automatically were introduced.
	Using a set of real-world datasets, we assessed the effectiveness of our automatic summary graph generation approach. 
	For future work, we plan to focus on the scalability of the proposed method in very large datasets.

	\begin{acknowledgements}
		The authors would like to thank Prof. Austin Melton for his invaluable help and his guidance during the study, Dr. Ruoming Jin and Dr. Viktor Lee for sharing RoleSim similarity measure.
		
	\end{acknowledgements}


	\bibliographystyle{spmpsci} 
	\bibliography{mybibfile}
	
\end{document}